\documentclass[%
superscriptaddress,
twocolumn,
 amsmath,amssymb,
 aps,
floatfix,
]{revtex4-1}

\usepackage{graphicx}
\usepackage{dcolumn}
\usepackage{bm}
\usepackage[utf8]{inputenc}
\usepackage[T1]{fontenc}
\usepackage{romannum}
\usepackage{xr}
\usepackage{color}
\usepackage{amsmath}
\usepackage[color=green!40]{todonotes}
\usepackage{soul}
\setstcolor{red}

\newcommand{\Tref}{T_{\rm{L}}}
\newcommand{\ci}{{\bm{c}_i}}
\newcommand{\vi}{{\bm{v}_i}}
\newcommand{\x}{\bm{x}}
\newcommand{\U}{\bm{u}}
\newcommand{\e}{e}
\newcommand{\dt}{\partial_t^{(1)}}
\newcommand{\dtt}{\partial_t^{(2)}}
\newcommand{\dalpha}{\partial_\alpha^{(1)}}
\newcommand{\dbeta}{\partial_\beta^{(1)}}
\newcommand{\dgamma}{\partial_\gamma^{(1)}}
\newcommand{\lp}{\left(}
\newcommand{\rp}{\right)}

\begin{document}

\title{Thermokinetic model of compressible multiphase flows}

\author{E. Reyhanian}
\affiliation{Department of Mechanical and Process Engineering, ETH Zurich, 8092 Zurich, Switzerland}
\author{B. Dorschner}
\affiliation{California Institute of Technology, Pasadena, CA 91125, USA}
\author{I. V. Karlin}\thanks{Corresponding author}
\email{karlin@lav.mavt.ethz.ch}
\affiliation{Department of Mechanical and Process Engineering, ETH Zurich, 8092 Zurich, Switzerland}
\date{\today}

\begin{abstract}
We present a novel approach to kinetic theory modeling enabling the simulation of a generic, real gas presented by its corresponding equation of state. The model is based on mass, momentum and energy conservation, and unlike the lattice Boltzmann models enables simulations of fluids exhibiting a liquid-vapor phase transition in both super- and subcritical states. In this new kinetic theory, an arbitrary equation of state can be introduced through rescaling the discrete particle velocities. Different benchmarks associated with real-gas thermodynamics illustrate that the proposed scheme can handle a wide range of compressible multiphase flows.
\end{abstract}

\maketitle
The lattice Boltzmann method (LBM)
	\cite{frisch1986lattice,mcnamara1988use,higuera1989lattice,qian1992lattice} is
	a modern approach to the simulation of complex flows. 
	LBM is a recast of fluid mechanics into a kinetic theory for the populations
	of designer particles $f_i(\bm{x}, t)$, with simple rules of propagation on a
	space-filling lattice formed by discrete speeds $\mathcal{C}=\{\bm{c}_i$, $i=1,\dots, Q\}$, in
	discrete-time $t$, and relaxation to a local equilibrium $f_i^{\rm eq}(\bm{x}, t)$  
	at the nodes $\bm{x}$. Classically, LBM features fluid dynamics as a fluctuation (subject to suitable nonlinearities) over a global thermodynamic reference state characterized by a lattice temperature $\Tref$ and global Galilean reference frame $\bm{u}=\bm{0}$. While this viewpoint led to successful LBM for incompressible flow, limitations are also apparent, in particular, for complex thermo-hydrodynamic processes for two-phase fluids. In order to address the entire spectrum of multiphase flows, from sub- to supersonic, involving sub- as well as supercritical processes, one needs a formulation of kinetic theory in a local rather than global thermodynamic reference frame \cite{PonD}.

In this paper, we present a new discrete-velocity kinetic theory of two-phase fluids as an extension of recently proposed "Particles on Demand for Kinetic Theory" method or "PonD" \cite{PonD}. Explicit realization of a local thermodynamic reference frame enables the simulation of demanding flow situations such as real-gas anomalous shock wave and droplet-shock interaction.

We follow \cite{PonD} and define discrete velocities
\begin{equation} \label{eq particle velocity}
\vi=\sqrt{\frac{p}{\rho \Tref}}\ci+\U,
\end{equation}
where $p(\bm{x},t)$ is the local thermodynamic pressure, $\rho(\bm{x},t)$ is the local density and $\Tref$ is a lattice reference temperature, a constant known for any set of speeds $\mathcal{C}$, and $\bm{u}(\bm{x},t)$ is the local flow velocity.
While our theory applies to a generic equation of state (EoS) $p$,  below we adopt the van der Waals (vdW) equation of state,
$ p = {\rho RT}/{(1-b\rho)} - a\rho ^2 $,
with critical values $\rho_{\rm cr} = 1/{3b}$, $T_{\rm cr} = 8a/27Rb$, $p_{\rm cr}=a/(27b^2)$; we set $a =2/49, b =2/21$ and $R=1$ in the simulations.

Local fields in (\ref{eq particle velocity}) are evaluated using two sets of populations, $f$ and $g$; the former 
maintain the density and momentum while the latter corresponds to energy conservation:
\begin{align}
\rho = \sum_{i=1}^{Q} f_i, \
\rho \U = \sum_{i=1}^{Q} f_i \vi , \
2\rho E = \sum_{i=1}^{Q} g_i,
\end{align}
where $E$ is the total energy per unit mass.
We comment that using the two-population approach is necessary for a generic equation of state. Indeed, only for the case of ideal, monatomic gas can the energy conservation be maintained by using the $f$-populations alone, where $2\rho E_{\rm K}=\rho u^2+D \rho T=\sum_{i=1}^{Q}f_iv_i^2$ and $D$ is the space dimension. It is only in this special case that one-population kinetic models can be justified while two populations are already needed if internal degrees of freedom are to be accounted for. For a generic real gas, the ideal gas may only result as a special limit of pressure approaching to zero where the internal energy is a function of temperature only. Therefore, there is no reason to define the temperature from ideal gas relation, and hence the total energy should be used instead.

For the sake of presentation, we shall at first neglect the interface energy and only consider $E = u^2/2 + \e$, where $\e=\e(s,v)$ is the local internal energy per unit of mass, $s$ is the entropy, $v=1/\rho$ the specific volume and temperature is defined by $T=(\partial \e/\partial s)_v$.

As in LBM, the proposed scheme is split into two parts; streaming and collision. Semi-Lagrangian advection \cite{PonD} is adopted in the streaming step:
\begin{align}
    f_i(\x,t) &=f_i(\x-\vi \delta t,t-\delta t), \label{eq f streaming}\\
    g_i(\x,t) &=g_i(\x-\vi \delta t,t-\delta t). \label{eq g streaming}
\end{align}

The collision step is then performed employing the Bhatnagar-Gross-Krook (BGK) model 
\begin{align} \label{eq f collision}
f_i^* (\x,t) &= f_i (\x,t) + \omega (f_i^{\rm eq} - f_i (\x,t)),\\
g_i^*(\bm{x},t) &= g_i(\bm{x},t) + \omega (g_i^{\rm eq} - g_i(\bm{x},t)) +\Tilde{g}_i \delta t \label{eq g collision},
\end{align}
where local equilibrium populations in the co-moving reference frame reduce to
\begin{align} 
f_i^{\rm eq}&=\rho W_i,\\
\label{eq geq}
		g_i^{eq}&=\rho W_i \left[2\e - {D}({p}/{\rho}) +  v_i^2\right],
\end{align}
where $W_i$ are lattice weights, which are known for any set $\mathcal{C}$. The equilibrium $g_i^{eq}$ is obtained by using Grad's distribution in the co-moving reference frame \cite{karlin2013consistent}. The correction term $\Tilde{g}_i \delta t$ on the R.H.S of Eq. (\ref{eq g collision}) is employed to impose the correct heat flux and takes the following form in the co-moving reference frame,
\begin{equation}
    \tilde{g}_i = {M_0}{W_i}{\lp 1+ \frac{\rho(\U\cdot\ci)^2}{2pT_{\rm L}} -\frac{\rho v_i^2}{2p} +\frac{D}{2} \rp}, \label{eq gtilde}
\end{equation}{}
where $M_0 = \sum{\tilde{g}_i} = 2\nabla\cdot(-\mu\nabla h+k\nabla T)$ is the correction in the energy equation, $\mu$ is the shear viscosity, $h = e + p/{\rho}$ is the enthalpy and $k$ is the conductivity which can be set independently.

We comment that the model kinetic equation treats the {\it total} energy as a local conservation by the collision term of $g$-populations (\ref{eq g collision}), whereas the $f$-population's collision conserves mass and momentum. The former conservation of the total local energy is different in nature from more microscopic yet phenomenological theories such as Vlasov-Enskog and similar kinetic equations where only the kinetic energy, 
$2\rho E_{\rm K}=D\rho RT + \rho u^2 $ is conserved by collisions and the "rest" of the energy is maintained non-locally  by the Vlasov mean-field term. 
As a result the heat flux does not automatically yield the well-known Fourier law but can readily be recovered by inclusion of $\tilde{g}$ 
(see appendix for further details).

Kinetic system (\ref{eq f streaming}-\ref{eq geq}) recovers the equations for density, flow velocity and temperature in the hydrodynamic limit as follows (see appendix for details):
	\begin{align}
&	D_t\rho=-\rho\nabla\cdot\U, \label{eq:density}\\
&	\rho D_t\U = -\nabla p -\nabla\cdot\bm{\tau},\label{eq:momentum} \\
&\rho C_v D_t T=-\bm{\tau}:\nabla\U- T \left(\frac{\partial p}{\partial T}\right)_v \nabla\cdot\U
-\nabla\cdot \bm{q}^{\rm neq},\label{eq:temperature}
	\end{align}

where $D_t={\partial}_t+\U\cdot\nabla$ is the material derivative,  $C_v=(\partial \e /\partial T)_v$ is the specific heat at constant volume, $\bm{q}^{\rm neq}=-k\bm{\nabla }T$ is the heat flux and the nonequilibrium stress tensor reads,

	\begin{align}
	\bm{\tau}= -\mu \lp \nabla\U+\nabla\U^{\dagger}-\frac{2}{D}(\nabla\cdot\bm{u})\bm{I}\rp-\eta(\nabla\cdot\U)\bm{I}.
	\end{align} 
The shear and bulk viscosity are, 
	\begin{align}
	\mu &= \lp \frac{1}{\omega}-\frac{1}{2} \rp p\delta t , \\
	\eta &= \lp \frac{1}{\omega}-\frac{1}{2} \rp \lp \frac{D+2}{D}-\frac{\rho \varsigma^2}{p} \rp p\delta t,
	\end{align}
respectively and $\varsigma=\sqrt{(\partial p/\partial \rho)_s}$ is the speed of sound. 
Note, as expected, the bulk viscosity vanishes in the limit of ideal monatomic gas, $p\to \rho RT$, $\varsigma^2\to (D+2)RT/D$.

\begin{figure}
    \centering
    \includegraphics[width=1.0\linewidth]{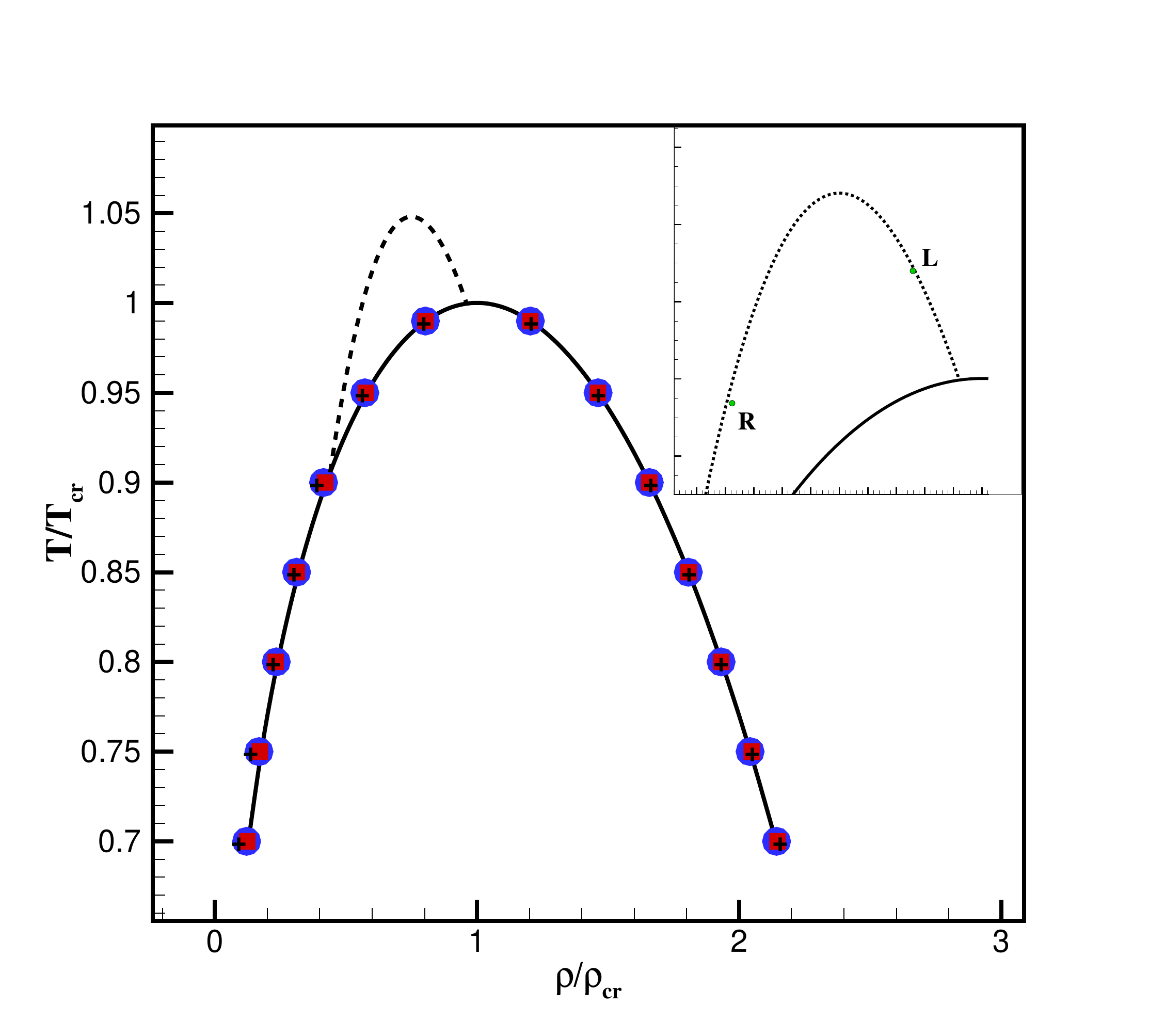}
    \caption{Coexistence curve of a van der Waals fluid. Solid line: Theory (Maxwell's equal area rule); Symbol: Simulation; Circle: $\delta = 1/3$; Square: $\delta = 2/3$; Cross: $\delta = 1/3$ advected with the speed $U=5U_{\rm ch}$, characteristic velocity $U_{\rm ch}=\sqrt{p_{\rm cr}/\rho_{\rm cr}}$. Dashed line: Zero line of fundamental derivative, $\Gamma = 0$, for $\delta = 0.0125$ (theory). Inset shows the left and right conditions in the simulation of the shock-tube problem in Fig.\ \ref{fig:4}. 
    }
    \label{fig:2}
\end{figure}

We proceed with the validation of the thermodynamical features of the proposed kinetic model.
The standard D2Q9 lattice $\mathcal{C}=\mathcal{C}_1\otimes\mathcal{C}_1$, where $\mathcal{C}_1=\{-1,0,1\}$ was used in all simulations.
For the van der Waals fluid, the internal energy is given by $\e = C_vT-a\rho$ and the specific heat at constant volume is $C_v = R/\delta$, where $0<\delta\leq2/3$ is a dimensionless parameter \cite{liu2018thermal, zhao2011admissible,colonna2006molecular}.
Fig.\ \ref{fig:2} demonstrates the independence of saturated liquid and vapor densities on the choice of $\delta$, also in a moving reference frame; results are in excellent agreement with Maxewll's equal-area rule.
Speed of sound is the next important benchmark;

It was measured by introducing a pressure disturbance $\delta p =10^{-3}$ in the liquid/gas phases (assumed at the saturation, and tracking the resulting shock front).
  
\begin{figure}
    \centering
    \includegraphics[width=1.0\linewidth]{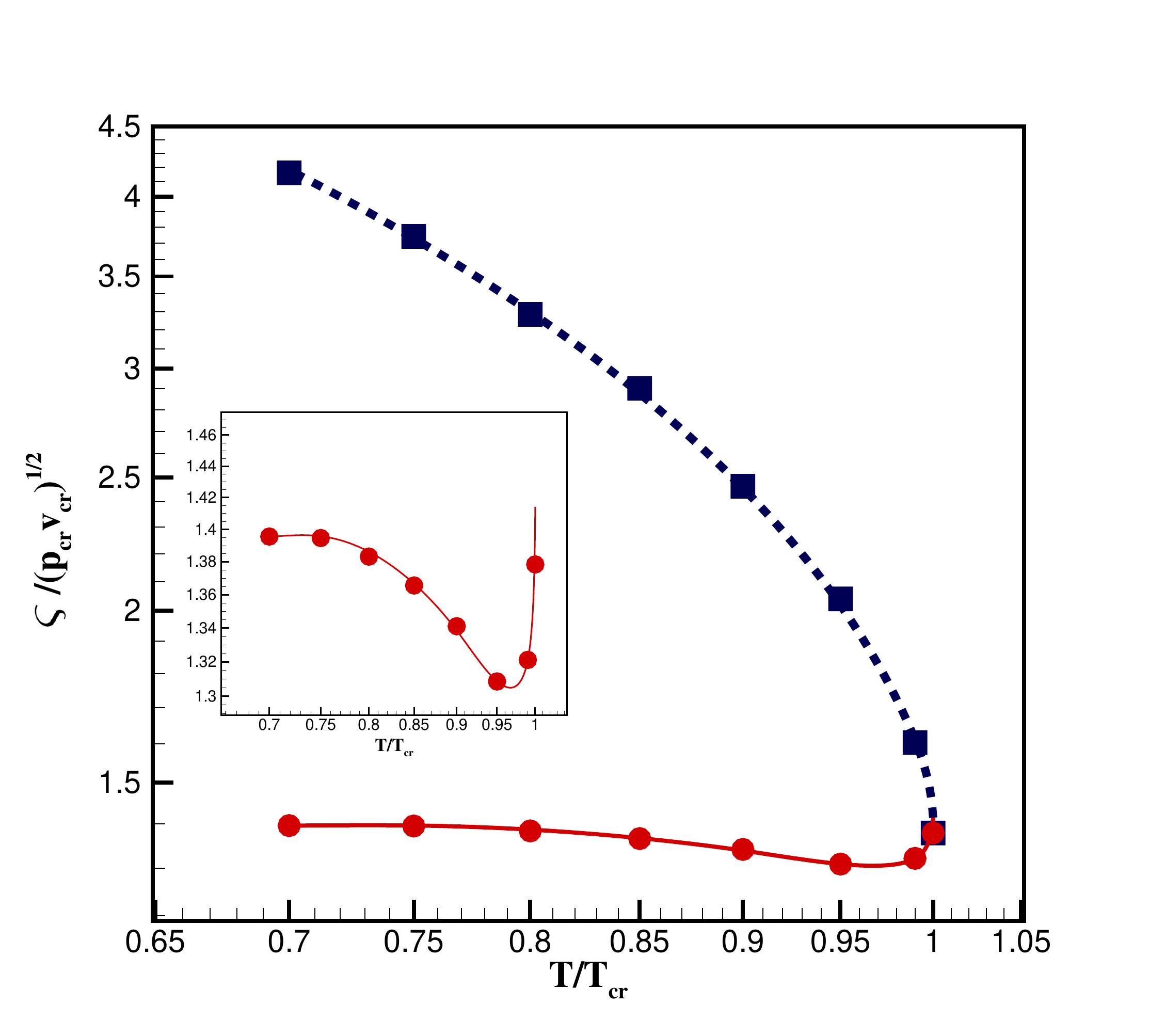}
    \caption{Reduced speed of sound in the van der Waals fluid with $\delta = 1/3$ as a function of the reduced temperature. Solid line: Saturated vapor (theory);  Dashed line: Saturated liquid (theory); Symbol: Simulation. Inset shows details of the speed of sound in the vapor near critical point.}
    \label{fig:1}
\end{figure}

Simulation results compare well with the theoretical prediction $\varsigma^2 = ({C_p}/{C_v})\lp{\partial p}/{\partial \rho}\rp_T$ in Fig.\ \ref{fig:1}. Note that the simulation predicts the speed of sound correctly at critical point $\varsigma_{\rm cr}=\sqrt{2p_{\rm cr}v_{\rm cr}}$.
The latter result is nontrivial because the finiteness of the speed of sound at critical point is maintained by the simultaneous divergence of $C_p$ and vanishing of the derivative $(\partial p/\partial \rho)_{T}$.

The inset in Fig.\ \ref{fig:1} demonstrates a non-monotonic behaviour of the speed of sound in the vapor phase for sufficiently large $C_v$ ($\delta=1/3$): a  decrease with the increase of the temperature up to $T/T_{\rm cr}=0.95$ followed by a sharp increase, matching the liquid line at the critical point.

\begin{figure}
    \centering
    \includegraphics[width=1.0\linewidth]{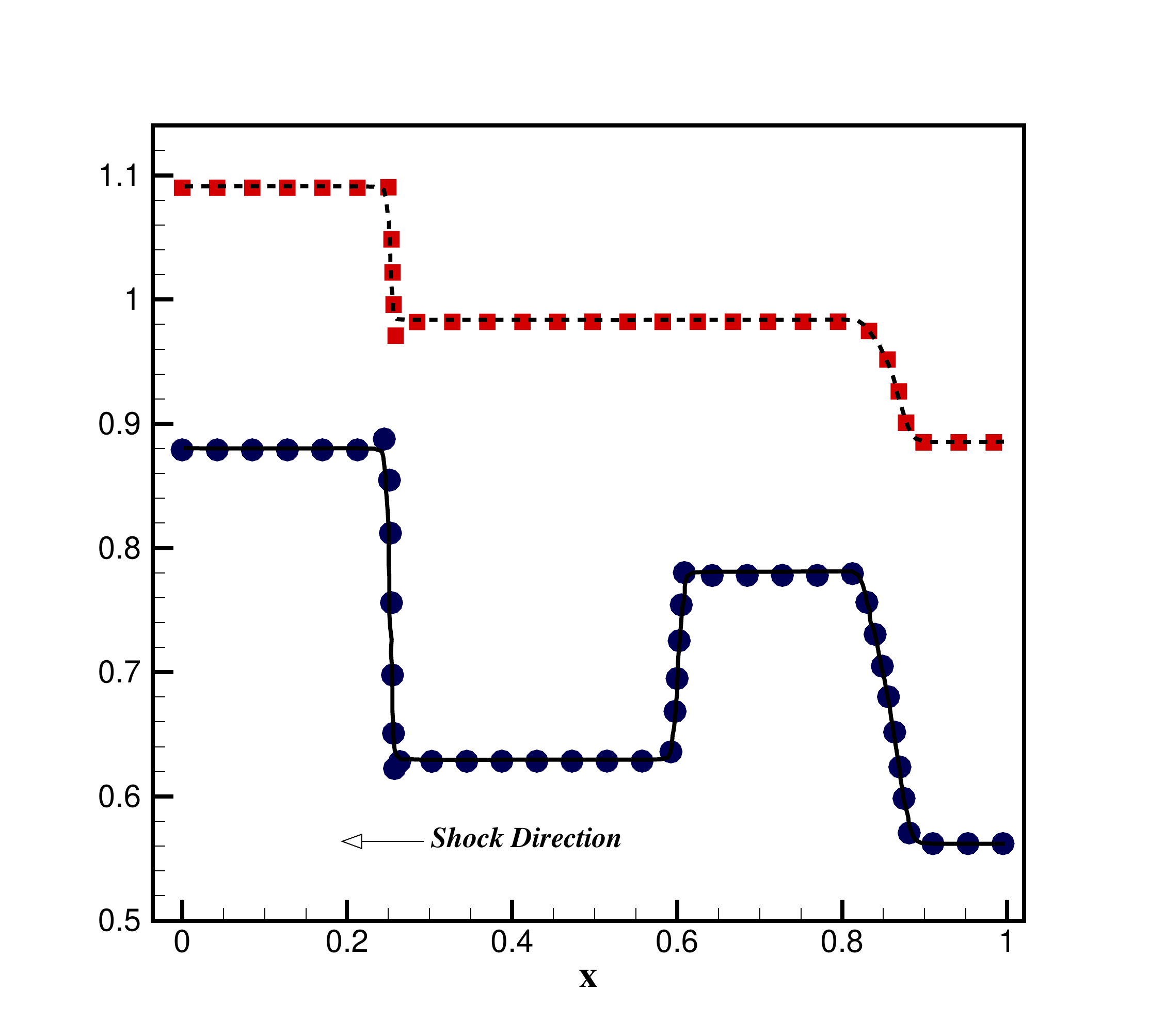}
    \caption{Simulation of an anomalous shock in the van der Waals fluid, $\delta = 0.0125$. Initial conditions: $(p_{\rm L},\rho_{\rm L}) = (1.09,0.879)$, $(p_{\rm R},\rho_{\rm R}) = (0.885,0.562)$ were applied to the left and right part of the tube. The snapshot is taken at time $t^\ast = (t/L)\sqrt{p_{\rm cr}/\rho_{\rm cr}}=0.45$, where $L$ is the length of the domain. 
     Line: Density \cite{guardone2002roe}, dashed: Pressure \cite{guardone2002roe}, symbols: Present.}
    \label{fig:4}
\end{figure}

As a final example, we consider the so-called anomalous shock-wave problem. For ideal gas, only compression shocks and rarefaction waves are observed. However, in the real-gas framework, it has been noted that there exist a region near the vapor saturation line where an anomalous behaviour may be observed; in terms of the existence of a rarefaction shock, travelling in the direction of the increase of pressure. Fluids demonstrating such a behaviour are known as Bethe-Zel'dovich-Thompson (BZT) fluids \cite{zhao2011admissible,bates1999some,guardone2002roe,zamfirescu2008admissibility}. 
The anomalous behaviour is characterized by the so-called fundamental derivative, $\Gamma = (v^{3} /{2\varsigma^2})(\partial^2 p/\partial v^2)_s$, $v=1/\rho$. An anomalous shock occurs when $\Gamma<0$ or when the $\Gamma=0$ boundary is crossed during the evolution of the flow and it can be modeled with the vdW fluid at large specific heat values (see dashed line and inset in Fig.\ \ref{fig:2}). 
Fig.\ \ref{fig:4} shows the snapshot of density and pressure profiles in the shock tube simulation when the left and right ends of the domain are in the $\Gamma<0$ domain (see inset in Fig.\ \ref{fig:2}). The anomalous rarefaction shock is clearly visible here, traveling toward the high pressure part together with a compression wave propagating to the right. 
The comparison between the present scheme and Ref. \cite{guardone2002roe} also shows good agreement.

\begin{figure}
    \centering
    \includegraphics[width=1.0\linewidth]{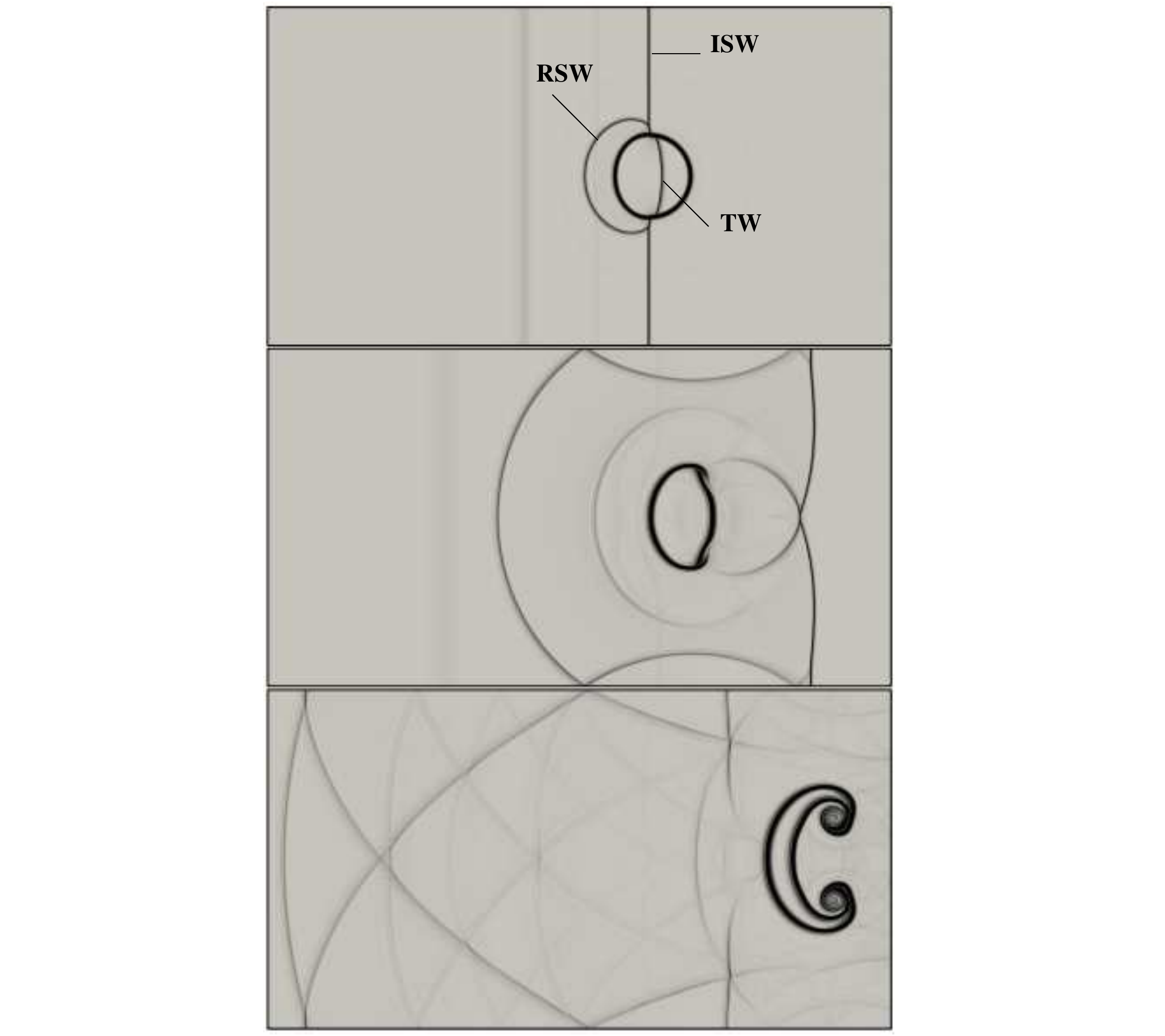}
    \caption{Schlieren images of the interaction of a shock wave at $\rm{Ma}=1.47$ and a droplet at different times using the present scheme. From top to bottom: $t^\ast=0.298$, $t^\ast=0.879$, $t^\ast=2.413$ }
    \label{fig:5}
\end{figure}

Up to now, our formulation was purely local, however in order to describe two-phase flows, we now extend the above local kinetic model to account for the non-local effects in the liquid-vapor interface. This is done by the following modifications:
The collision step for the $f$-populations (\ref{eq f collision}) becomes augmented with a source (forcing) term $S_i$,
\begin{align}
f_i^* (\x,t) &= f_i (\x,t) + \omega (\rho W_i - f_i (\x,t))+S_i,
\end{align}
where
\begin{align}
S_i=\mathcal{G}_{\bm{u}+\delta\bm{u}}^{\bm{u}}[\rho W_i] -\rho W_i. \label{eq S-i}
\end{align}
$\mathcal{G}$ is the transformation matrix (see \cite{PonD} and Supplemetal Material) and $\delta \U = \mathbf{F}/\rho \delta t$ is the change of the local flow velocity due to the force $\mathbf{F}=\nabla \cdot \bm{K}$, where
\begin{equation} \label{eq korteweg}
\bm{K} =-\kappa(\rho \nabla\cdot\nabla \rho + \frac{1}{2} |\nabla \rho|^2)\bm{I} + \kappa\nabla\rho \otimes \nabla\rho,
\end{equation}
is the Kortwewg stress \cite{PhysRevLett.114.174502}. The first term on the R.H.S of equation (\ref{eq S-i}) denotes the transformation of equilibrium populations residing at the reference frame $"\U+\delta\U"$ to the reference frame $"\U"$ [See appendix]. Having included the source term $S_i$, the actual fluid velocity is now shifted to $\hat{\U} = \U + \delta\U/2$ where $\U=1/\rho \sum f_i\vi$.
In a similar manner, one can incorporate a source term $\phi_i$ for the g-population to recover the effect of the Korteweg stress in the energy equation such that $\sum \phi_i=2\hat{\U} \cdot \nabla \cdot \bm{K}$ is satisfied. Hence the post-collision $g$-populations (\ref{eq g collision}) is recast in the form,
\begin{align}
g_i^* (\x,t) = g_i (\x,t) + \omega (g_i^{eq} - g_i (\x,t)) + \Tilde{g}_i\delta t +{\phi_i}\delta t,
\end{align}
The equilibrium (\ref{eq geq}) is now extended to take into account the effect of the force,
\begin{equation}
    g_i^{eq}=W_i \left[ 2\rho E+ \bm{M} \cdot (\vi-\U) + \bm{N}:(\vi\vi-({p}/{\rho})\bm{I}-\U\U) \right],
    \label{eq new geq}
\end{equation}
\begin{equation}
    \bm{M}=\frac{\delta t}{(p/\rho)}\left[ \bm{F}H-\lp \hat{\U}\bm{F}+\bm{F}\hat{\U}+\delta t \bm{F}\bm{F}\frac{E}{2p}\rp\cdot\bm{\U} \right],
\end{equation}
\begin{equation}
    \bm{N}=\rho\bm{I}+ \frac{\delta t}{(2p/\rho)}\left[ \hat{\U}\bm{F}+\bm{F}\hat{\U}+\delta t \bm{F}\bm{F}\frac{E}{2p}\right],
\end{equation}
where $H=\e+\hat{u}^2/2+p/\rho$ is the total enthalpy. It should be noted that in the absence of the force, the new equilibrium (\ref{eq new geq}) simplifies to Eq. (\ref{eq geq}). With the mentioned changes, the momentum equation (\ref{eq:momentum}) is modified as,
\begin{align}\label{eq:momentum_K}
    	\rho D_t\hat{\U} = -\nabla p -\nabla\cdot\bm{\tau} -\nabla\cdot \bm{K},
\end{align}
while the density (\ref{eq:density}) and the temperature (\ref{eq:temperature}) equations stay intact with the only difference that $\U$ is replaced by $\hat{\U}$.

We conclude this paper with a simulation of the interaction of a water column with a planar shock wave. To this end, a planar shock wave was generated initially separating the post-shock part and the saturated vapor. 
The droplet is present in the downstream of the shock where it is initially in equilibrium with its vapor at the temperature $T/T_{\rm cr}=0.9$. 

In order to compare with the experiment \cite{xiang2017numerical, igra2003experimental}, we use the reduced time
 $ t^\ast = t (u_{g}/d_0) \sqrt{\rho_{\rm g}/\rho_{\rm l}}(\varsigma_{l}/\varsigma_{g})$ where $\varsigma_{l,g}$ is the  speed of sound in the liquid and the post-shock gas, $\rho_{l,g}$ are corresponding densities, $d_0$ is the initial diameter of the liquid column and $u_g$ is the flow speed upstream.

\begin{figure}
    \centering
    \includegraphics[width=1.0\linewidth]{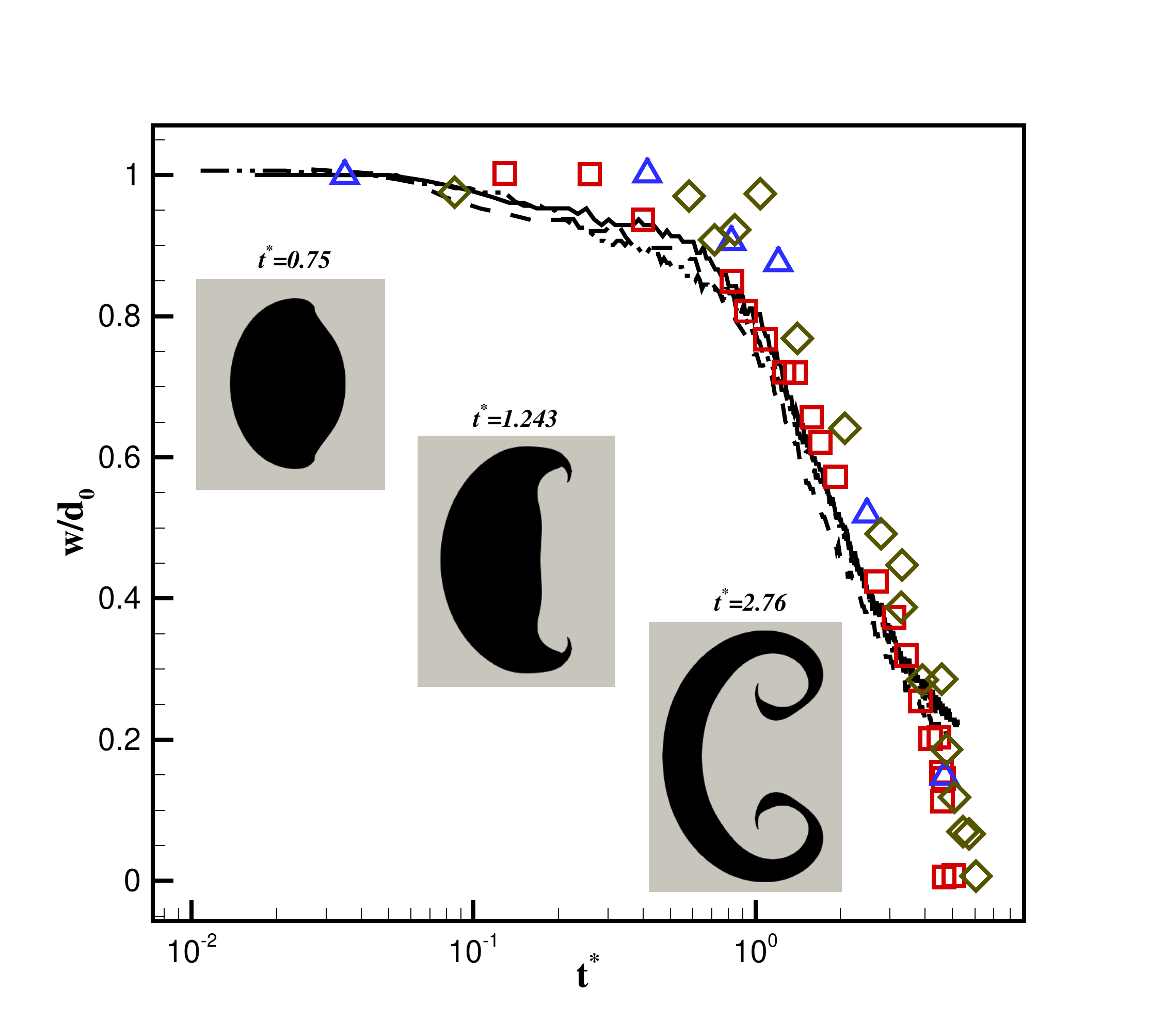}
    \caption{2D droplet center-line width evolution. Lines: Present scheme, solid line: $M_s=1.18$, long dashed: $M_s=1.33$, dashed: $M_s=1.47$. Symbol: Experiment \cite{igra2001study}, square: $M_s=1.18$, delta: $M_s=1.30$, diamond: $M_s=1.47$.}
    \label{fig:6}
\end{figure}

The evolution of the water-column's shape is illustrated in Fig. \ref{fig:5} in terms of Schlieren images. After the incident shock wave (ISW) reaches the upstream interface of the droplet, one can see that a left-propagating reflected shock wave (RSW) as well as the transmitted wave (TW) are generated. The TW quickly travels the width of the droplet since the speed of sound is significantly higher in the liquid phase than in the vapor. These are typical waves generated in the early stages upon the impingement of a shock wave on a droplet as reported in experiments and other numerical simulations \cite{sembian2016plane,xiang2017numerical}.\\
At later stages, the droplet starts to flattening in flow direction and expanding in radial direction. 
Furthermore, two vortices are formed near the equator, which are the result of the shear forces and the flow separation behind the droplet.

Fig. \ref{fig:6} represents the quantitative assessment of the simulation where the width of the droplet was measured with respect to its center-line and compared to the experimental results from \cite{igra2001study}. In addition the interface of the droplet is depicted at three different times. The results show an excellent agreement with the experiment showing that the deformation of the droplet was accurately captured by the proposed scheme.

In conclusion, the main novelty of our approach is to introduce local thermodynamics by a simple velocity rescaling whereas the rest follow automatically provided the local conservation laws are correctly taken into account by model collision. The restriction of the flow velocity in LBM to small values limits its application in high-velocity flows leading the compressible-multiphase flow regime an uncharted field in the context of LBM. The Galilean-invariant nature of this method removes this barrier, on account of error-free collision. Real-gas EOS is naturally introduced through the discrete particle velocities, providing full thermodynamic-consistency. Furthermore, the proposed model conserves total energy, thereby enabling simulations such as anomalous shock-waves in a real gas, isentropic speed of sound in a liquid-vapor system and interaction of a shock wave with a water-column.

The authors would like to thank Dr. Fabian Bösch for the helpful discussions.

This work was supported by the European Research Council (ERC) Advanced Grant No. 834763-PonD and the SNF Grants No. 200021-172640 (E.R.) and No. P2EZP2\_178436 (B.D.). Computational resources at the Swiss National Super Computing Center (CSCS) were provided under Grant No. s897.
\appendix
\section{THERMODYNAMICS}
\paragraph{Internal energy}
To capture the correct thermodynamics of the real gas, it is essential to use the correct thermodynamical relations. Real gases differ from ideal gas in the sense that the internal energy and enthalpy are no longer the function of temperature only. Instead, they depend on two thermodynamical parameters namely, $\e=\e(T,v)$ and $h=h(T,p)$ where $v$ is the specific volume of the fluid. Therefore the internal energy of a real gas can be described as
\begin{equation}
    d\e = C_vdT+\lp\frac{\partial \e}{\partial v}\rp_Tdv
    \label{eq SM real-gas internal energy}
\end{equation}
Where $C_v$ is the specific heat at constant volume which is considered as constant in the "polytropic" assumption \cite{zhao2011admissible, guardone2002roe,colonna2006molecular}. To compute the non-tangible part of Eq. (\ref{eq SM real-gas internal energy}), the well-known thermodynamical relation $Tds=d\e+pdv$ is put to use
\begin{equation}
    T\lp\frac{\partial s}{\partial v}\rp_T = \lp\frac{\partial \e}{\partial v}\rp_T+p
\end{equation}
Using the famous Maxwell relation $\lp\frac{\partial s}{\partial v}\rp_T = \lp\frac{\partial p}{\partial T}\rp_v$ leads to the differential equation for real gas internal energy
\begin{equation}
    d\e=C_v dT + \left[T\lp\frac{\partial p}{\partial T}\rp_v - p\right]dv
    \label{eq SM real-gas internal energy final}
    \end{equation}
\paragraph{Calculation of $C_p$}
Specific heat capacity at constant pressure is defined as
\begin{equation}
    C_p = \lp\frac{\partial h}{\partial T}\rp_p
    = \lp\frac{\partial \e}{\partial T}\rp_p + 
    p\lp\frac{\partial v}{\partial T}\rp_p
    \label{eq SM Cp definition}
\end{equation}
where  the expression of enthalpy $h = \e + pv$ is used. According to Eq. (\ref{eq SM real-gas internal energy final}) one can write
\begin{equation}
    \lp\frac{\partial \e}{\partial T}\rp_p
    = C_v +  \left[T\lp\frac{\partial p}{\partial T}\rp_v - p\right]\lp\frac{\partial v}{\partial T}\rp_p
\end{equation}
Finally, the expression for the $C_p$ of a real-gas is obtained as
\begin{equation}
    C_p = C_v + T\lp\frac{\partial p}{\partial T}\rp_v
    \lp\frac{\partial v}{\partial T}\rp_p
    \label{eq SM Cp expression}
\end{equation}
\paragraph{Speed of sound}
The theoretical description of speed of sound in a real-gas originates from its definition
\begin{equation}
  \varsigma^2 = \lp\frac{\partial p}{\partial \rho}\rp_s
    =-v^2 \lp\frac{\partial p}{\partial v}\rp_s
    \label{eq SM cs2 definition}
\end{equation}
Using the well-known cyclic relation, we obtain
\begin{equation}
    \lp\frac{\partial p}{\partial v}\rp_s = 
    -\lp\frac{\partial p}{\partial s}\rp_v \div
    \lp\frac{\partial v}{\partial s}\rp_p
    \label{eq SM cyclic1}
\end{equation}
The terms on the R.H.S of Eq. (\ref{eq SM cyclic1}) can be substituted by their counterparts using the Maxwell relations
\begin{align}
    -\lp\frac{\partial p}{\partial s}\rp_v &= 
    \lp\frac{\partial T}{\partial v}\rp_s
    \label{eq SM Maxwell1}\\
    \lp\frac{\partial v}{\partial s}\rp_p &= 
    \lp\frac{\partial T}{\partial p}\rp_s
    \label{eq SM Maxwell2}
\end{align}
Applying the cyclic relation to the R.H.S of Eqs. (\ref{eq SM Maxwell1}) and (\ref{eq SM Maxwell2}) leads to
\begin{align}
    -\lp\frac{\partial T}{\partial v}\rp_s &= 
    \frac{T}{C_v}\lp\frac{\partial p}{\partial T}\rp_v
    \\
    \lp\frac{\partial T}{\partial p}\rp_s &= 
    \frac{T}{C_p}\lp\frac{\partial v}{\partial T}\rp_p
\end{align}
Finally the expression for speed of sound is obtained as
\begin{equation}
\varsigma^2 = \frac{C_p}{C_v}\lp\frac{\partial p}{\partial \rho}\rp_T
\end{equation}
By using Eq. (\ref{eq SM Cp expression}), we get the following thermodynamical relation for speed of sound in a real gas medium
\begin{equation} \label{eq SM speed of sound formula}
\varsigma =\sqrt{ \lp\frac{\partial p}{\partial \rho}\rp_T + \frac{T}{\rho^2 c_v}\lp\frac{\partial p}{\partial T}\rp_\rho ^2 }
\end{equation}
Which implies that the isentropic speed of sound is always higher than that in isothermal condition.
For the van der Waals EoS, one can simply derive the following expressions,
\begin{align}
 e_{vdw}&=C_v T-a\rho,\label{eq SM vdW internal energy}\\
 {C_p}_{vdw}&=C_v+\frac{R^2T}{RT-2a\rho(1-b\rho)^2},\\
 {\varsigma}_{vdw}&=\sqrt{\frac{RT}{(1-b\rho)^2}(1+\delta)-2a\rho},
\end{align}
\section{EQUILIBRIUM}
We consider the standard nine-velocity model, the \textit{D2Q9} lattice. The discrete speeds are constructed as atensor product of two one-dimensional peculiar speeds,
$c_i = i$, where $i = 0, \pm 1$. Discrete speeds in two-dimensions are
\begin{equation}
    \mathbf{c}_{(i,j)} = (c_i,c_j)^\dagger,
    \label{eq SM peculiar velocities}
\end{equation}
where we have introduced two-dimensional indices in order to reflect the Cartesian frame instead of a more common single subscript. Thus, the discrete velocities are defined as
\begin{equation}
    \mathbf{v}_{(i,j)} =\sqrt{\theta}(c_i,c_j)^\dagger+(u_x,u_y)^\dagger
    \label{eq SM particle velocities}
\end{equation}
with reduced temperature $\theta = p/(\rho T_L)$ and lattice temperature $T_L = 1/3$. Populations are labeled with two indices, $f_{(i,j)}$, corresponding to their respective velocities (\ref{eq SM particle velocities}). The local equilibrium populations are now conveniently expressed as the product of one-dimensional
weights
\begin{equation}
    f^{eq}_{(i,j)} =\rho W_{(i,j)}=\rho W_i W_j
    \label{eq SM feq}
\end{equation}
Where
\begin{equation}
    W_i = \Big\{ \begin{tabular}{ll}
        2/3, & for i=0  \\
        1/6, & otherwise
    \end{tabular}
    \label{eq SM weights}
\end{equation}
While the equilibrium populations are constant up to the proportionality to density, their moments
\begin{equation}
        M_{mn}^{eq}=\rho \sum_{\substack{(i,j)}}W_i W_j(\sqrt{\theta} c_i+u_x)^m(\sqrt{\theta} c_j+u_y)^n,
\end{equation}
recover the pertinent nine Maxwell-Boltzmann moments up to the fourth order, $0 \leqslant m \leqslant 2$, $0 \leqslant n \leqslant 2$, $m+n \leqslant 4$, without error for any velocity.
\section{TRANSFER MATRIX}
Populations $f_{(i,j)}^\lambda$ measured in the gauge $\lambda$, can be represented as linear combinations of nine linearly independent moments,
\begin{equation}
    M^\lambda=(M_{00}^\lambda,M_{10}^\lambda,M_{01}^\lambda,M_{11}^\lambda,M_{20}^\lambda,M_{02}^\lambda,M_{21}^\lambda,M_{12}^\lambda,M_{22}^\lambda)^\dagger
    \label{eq SM Moments}
\end{equation}
Where
\begin{equation}
        M_{mn}=\sum_{\substack{(i,j)}}f_{(i,j)}^\lambda (\sqrt{\theta} c_i+u_x)^m(\sqrt{\theta} c_j+u_y)^n,
\end{equation}
and $\mathcal{M}^\lambda$ is the $Q \times Q$
matrix of the linear map between populations and moments,
\begin{equation}
    \mathcal{M}_\lambda f^\lambda = M^\lambda
\end{equation}
Moments are invariant with respect to the gauge,
\begin{equation}
    \mathcal{M}_{\lambda^\prime} f^{\lambda^\prime} = \mathcal{M}_\lambda f^\lambda
    \label{eq SM invariant moment}
\end{equation}
This implies that the populations are transformed from one gauge to another with the transfer matrix $\mathcal{G}_\lambda^{\lambda^\prime}$,
\begin{equation}
    f^{\lambda^\prime}=\mathcal{G}_\lambda^{\lambda^\prime}f^\lambda=\mathcal{M}_{\lambda^\prime}^{-1}\mathcal{M}_{\lambda}f^\lambda
\end{equation}
The transfer from gauge $\lambda$ to $\lambda ^\prime$ can also be written in the following explicit form,
\begin{equation}
        f_{kl}^{\lambda^\prime}=w(k)w(l) \sum_{\substack{(i,j)}}g_x(i,k)g_y(j,l)f_{(i,j)}^\lambda,
        \label{eq SM explicit form}
\end{equation}
Where
\begin{align}
   g_\xi(i,j) &=A_\xi^2(i)-B_\xi(i,j),\\
   A_\xi(i) &=\left(u^\prime_\xi-u_\xi \right)/\sqrt{3}-i\sqrt{p/\rho},\\
    B_\xi(i,j) &= \Big\{ \begin{tabular}{cl}
    $p^ {\prime} / \rho ^ {\prime}$, & for j=0,  \\
    $j\sqrt{p^\prime/\rho^\prime}A_\xi(i)$,  & otherwise,
    \end{tabular}\\
    w(i) &= \Big\{
    \begin{tabular}{cl}
    $1/(\frac{p^\prime}{\rho^\prime})$,& for i=0,  \\
    $-1/2(\frac{p^\prime}{\rho^\prime})$, & otherwise.
   \end{tabular}
\end{align}
Formula (\ref{eq SM explicit form}) only involves evaluation of a dot-product as opposed to numerically solving the linear system (\ref{eq SM invariant moment}).
\section{RECONSTRUCTION}
An equidistant rectilinear mesh with $\Delta x = 1$ is used for all simulations. Populations at off-grid locations are reconstructed using 3\textsuperscript{rd}-order polynomial interpolation,
\begin{equation}
        \tilde{f}^{\lambda}_{(i,j)}(\x,t)= \sum_{\substack{0\leq m \leq 3 \\ 0\leq n \leq 3}}a_{mn}\lp\x\rp {f}^{\lambda}_{(i,j)}((x_0+n,y_0+m),t),
\end{equation}
where the populations at integer collocation points $(x_0+n,y_0+m)$ are transformed to gauge $\lambda$ using Eq. (\ref{eq SM explicit form}) and $a_{mn}$ are standard Lagrange polynomials,
\begin{equation}
        a_{mn}(\x)= \prod_{\substack{0\leq k \leq 3 \\ k\ne n}}\frac{(x-x_0)-k}{n-k}
        \prod_{\substack{0\leq l \leq 3 \\ l\ne m}}\frac{(y-y_0)-l}{m-l},
\end{equation}
with respect to reference coordinate,
\begin{equation}
    \x_0=\lp \lfloor x \rfloor -1,\lfloor y \rfloor -1 \rp,
\end{equation}
where the operation $\lfloor \phi \rfloor$ rounds down to the largest integer value not greater than $\phi$.

\section{FURTHER SIMULATIONS}
\subsection{SPEED OF SOUND}

\begin{figure}
    \centering
    \includegraphics[width=8.6cm]{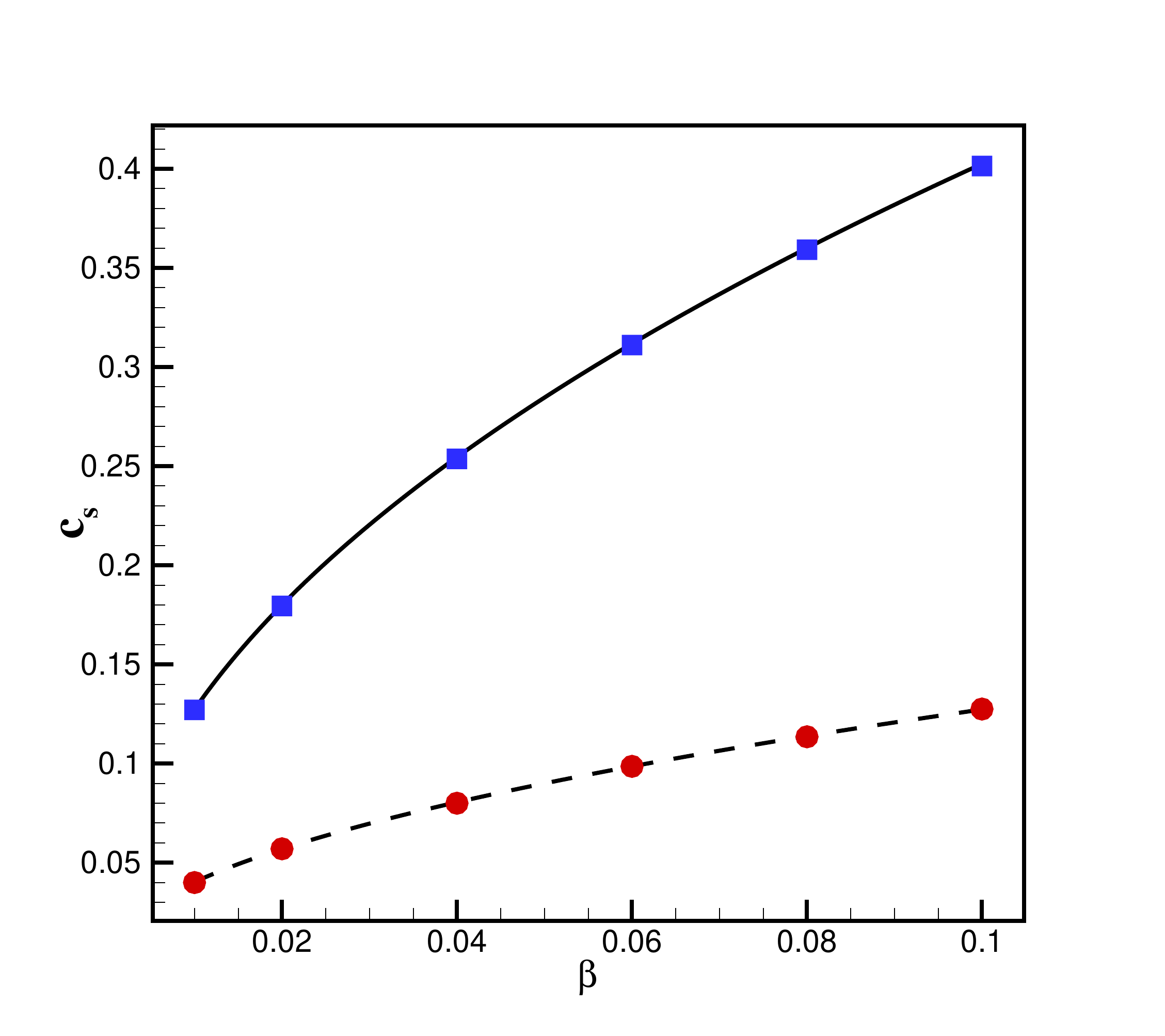}
    \caption{Speed of sound in a simplified EOS near the critical point. line: liquid (theory), dashed: vapor (theory), symbols: present. Values of speed of sound are preseted in lattice units.}
    \label{fig SM simplified EOS_Cs}
\end{figure}

\begin{figure}
    \centering
    \includegraphics[width=8.6cm]{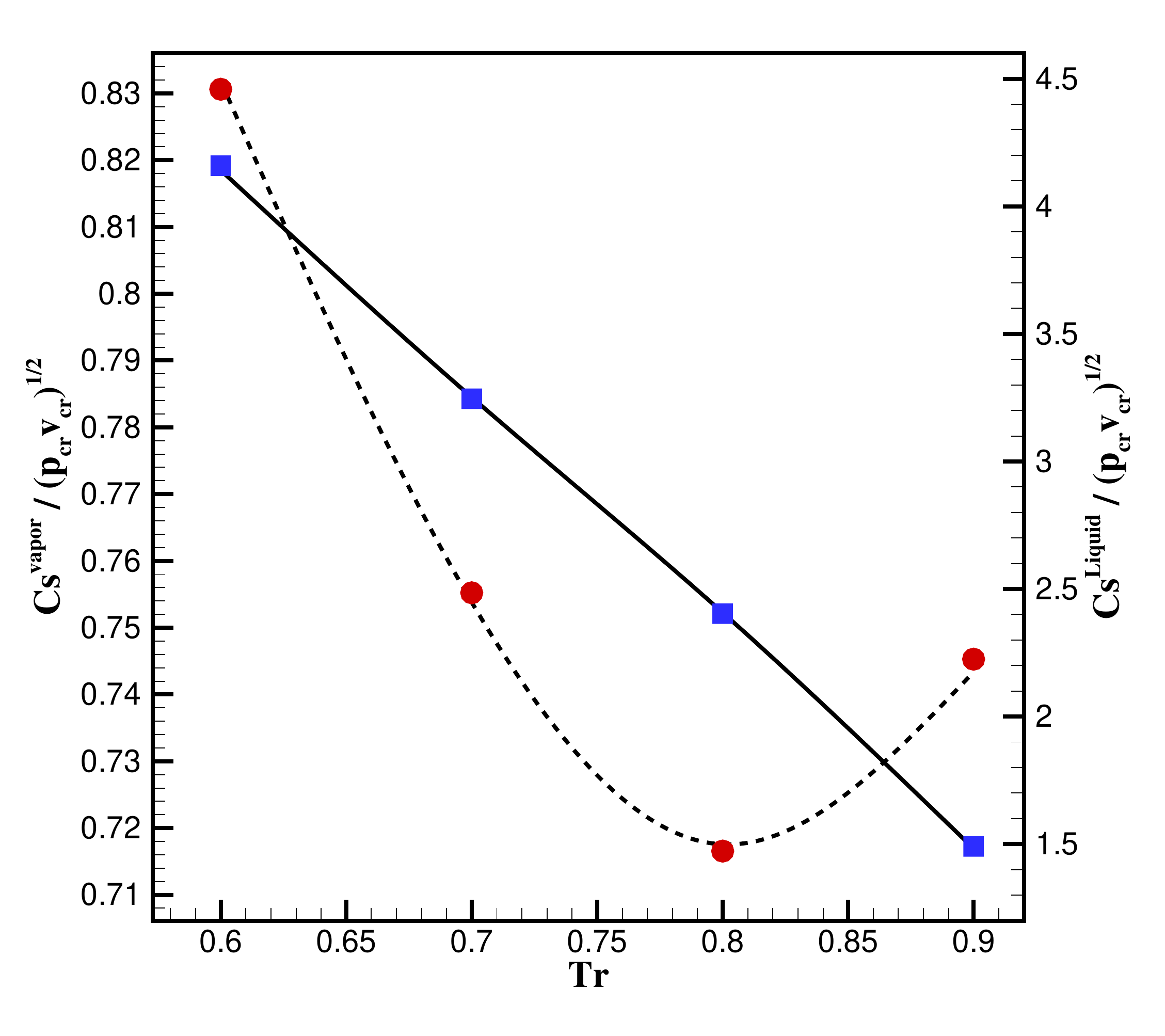}
    \caption{Isothermal speed of sound in a van der Waals fluid. line: liquid (theory), dashed: vapor (theory), symbols: present.}
    \label{fig SM Vdw isothermal Cs}
\end{figure}

First, we consider the simplified EOS near the critical point, where the free-energy takes the following form \cite{jamet2001second}
\begin{equation} \label{eq SM Free Energy formula}
E_0 = \beta (\rho-\rho_l^{sat})(\rho-\rho_v^{sat}),
\end{equation}
where $\beta$ is a function of temperature and ${\rho_l}^{sat}$ and ${\rho_v}^{sat} $ are the saturated densities of the liquid and vapor phase, respectively \cite{jamet2001second}. The thermodynamic pressure can now be obtained as $p=\rho \partial E_0 / \partial \rho -E_0$. In this special case, the isentropic speed of sound coincides with the isothermal one since the partial derivation of pressure to temperature at saturated liquid and vapor densities is simply zero. Hence, the speed of sound in the simplified EOS at saturated conditions reads
\begin{align}
\varsigma |_{\rho_l^{sat}} &= (\rho_l^{sat}-\rho_v^{sat})\sqrt{2\beta \rho_l^{sat}}, \label{eq SM Cs rhol}\\
\varsigma |_{\rho_v^{sat}} &= (\rho_l^{sat}-\rho_v^{sat})\sqrt{2\beta \rho_v^{sat}}, \label{eq SM Cs rhov}
\end{align}
choosing the density ratio as $\rho_l/\rho_v=10$, the measured value of speed of sound in the simulations are compared with theoretical expressions in Fig.\ref{fig SM simplified EOS_Cs}. The results for both liquid and gas phases show excellent agreement.
\\
The isothermal simulation is repeated with the same setup for the van der Waals fluid and Fig. \ref{fig SM Vdw isothermal Cs} shows excellent agreement between theory and simulation.
\subsection{COEXISTENCE CURVE}
\begin{figure}
    \centering
    \includegraphics[width=8.6cm]{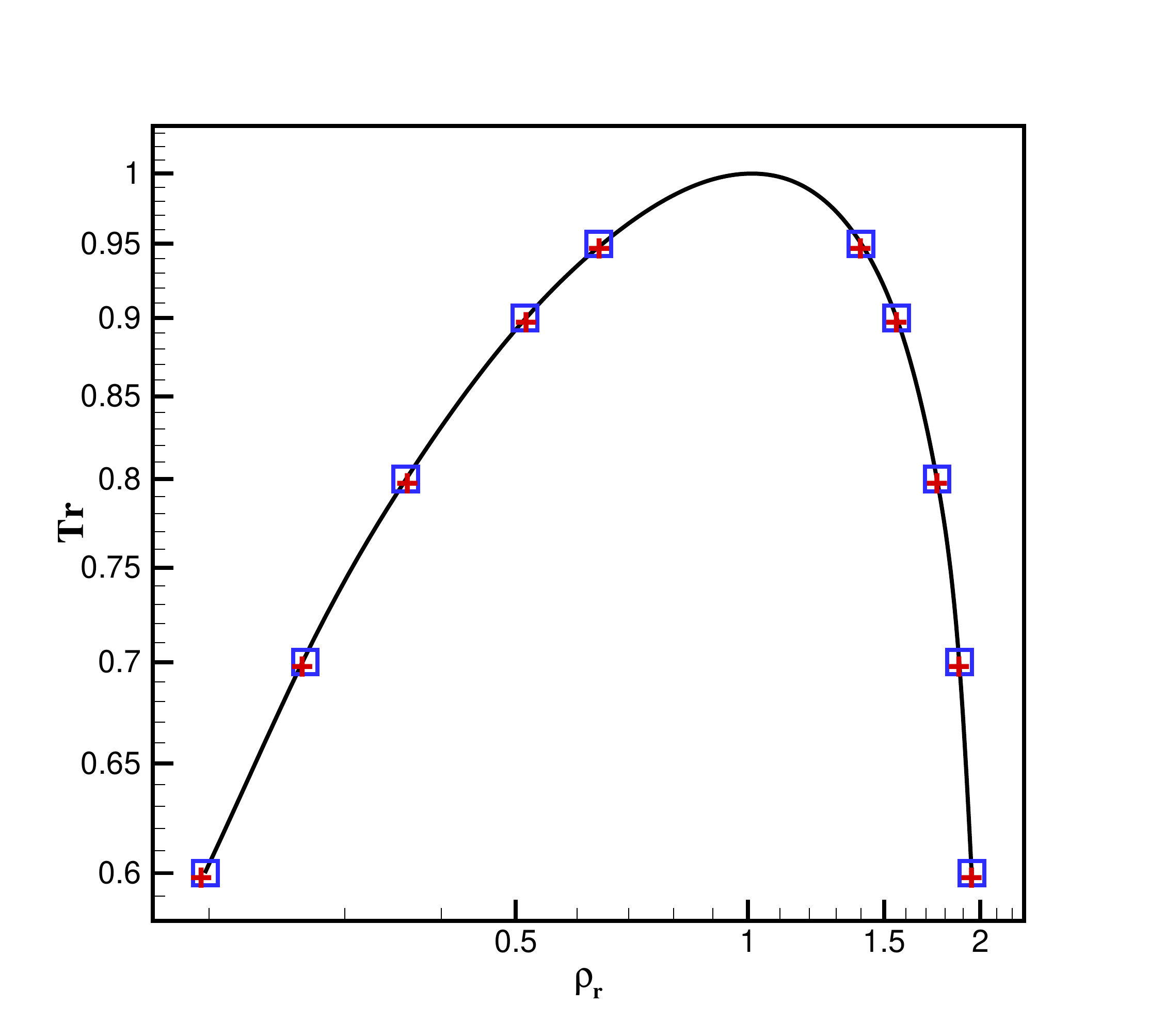}
    \caption{Isothermal simulation of the coexistence curve employing the Dietrici EOS. line: Maxwell, square: Present, cross: LBM.}
    \label{fig SM Dieterici coex}
\end{figure}

In this part, isothermal simulation of the coexistence curve is conducted with the Dieterici EOS
\begin{equation} \label{eq SM Dieterici}
p=\frac{\rho RT}{1-b\rho}exp(-\frac{a\rho}{RT})
\end{equation}
Introducing the critical properties as $\rho_{cr}=1/(2b)$, $T_{cr}=a/(4Rb)$ and $p_{cr}=a/(2eb)^2$ \cite{polishuk2004phase}, the reduced form of this EOS reads
\begin{equation} \label{eq SM Dieterici reduced}
p_r=\frac{\rho_r T_r e^2}{2-\rho_r}exp\lp-\frac{2\rho_r}{T_r}\rp
\end{equation}
where the reduced quantities are scaled to their corresponding critical values. Note that here $e=exp(1)$ is the Euler's number and it shall not be confused with internal energy. Simulation results are illustrated in Fig. \ref{fig SM Dieterici coex} and compared to predicted values by Maxwell construction.
\subsection{SURFACE TENSION}
\begin{figure}
    \centering
    \includegraphics[width=8.6cm]{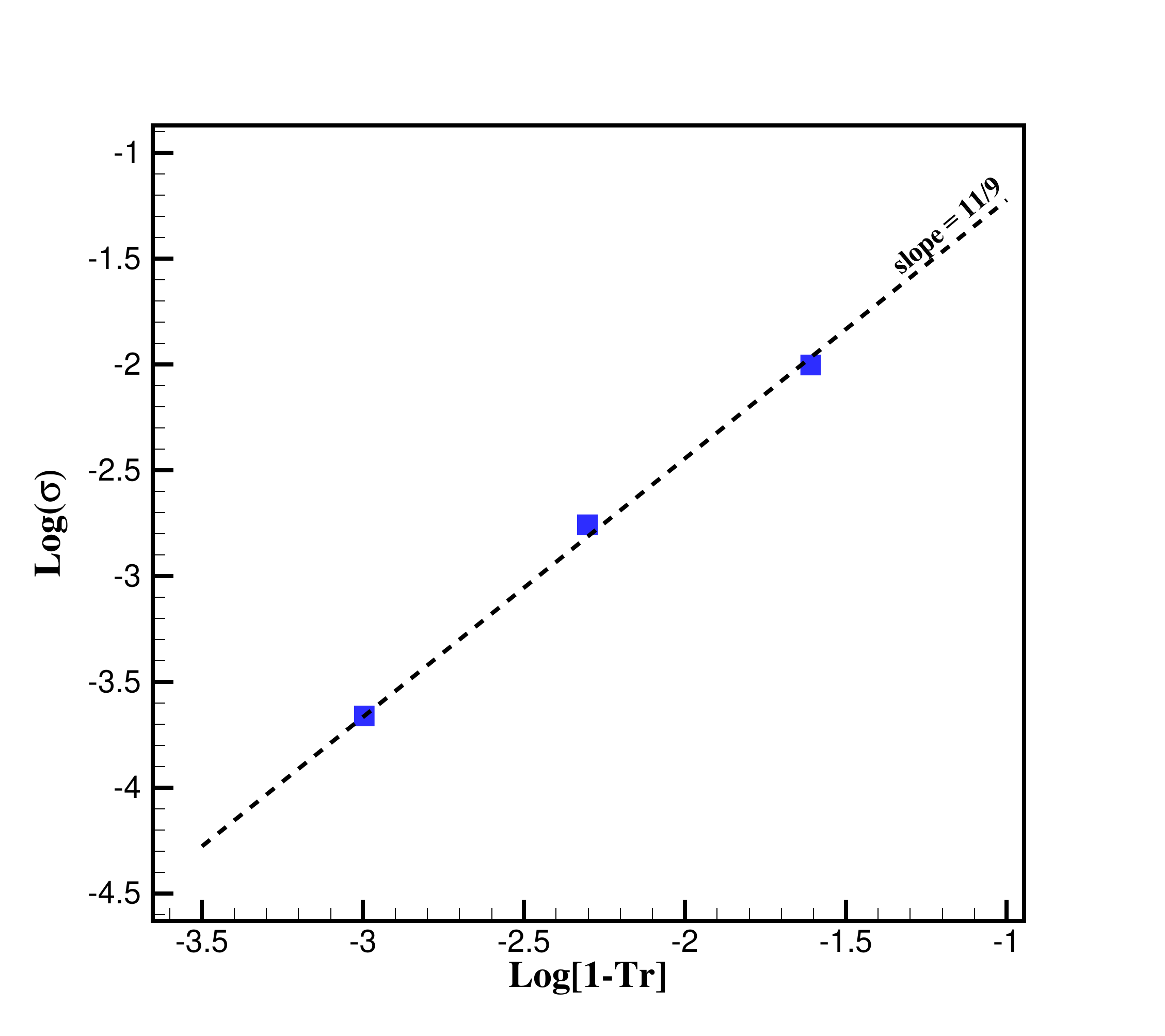}
    \caption{Temperature dependency of surface tension. square: present, dashed: empirical slope predicted by Eq. (\ref{eq SM :7})}
    \label{fig SM Dieterici surface tension}
\end{figure}

Continuing with the Dieterici EOS, the temperature dependency of the surface tension is evaluated. Surface tension of a liquid is a decreasing function of temperature and it vanishes when the system approaches the critical point. According to the Guggenheim$-$Katayama relation, this behavior can be formulated by \cite{guggenheim1945principle}
\begin{equation} \label{eq SM :7}
\sigma = \sigma_0 (1-T_r)^{11/9}
\end{equation}
Where $\sigma_0$ is a constant. The values of the surface tension is obtained by the Laplace law at each temperature simulating a 2D droplet. Fig. \ref{fig SM Dieterici surface tension} shows that there is good agreement between simulation results and predicted values by Eq. (\ref{eq SM :7}). 
\subsection{RANKINE-HUGONIOT CONDITIONS}
The shock front traveling with speed $s$ separates the domain into two parts; $\Omega_1$ the post-shock part and $\Omega_0$ the pre-shock part. Shock waves must satisfy the Rankine-Hugoniot relations
\begin{align}
    -s\langle M \rangle& + \langle N \rangle=0,\\
    M=\left[
    \begin{tabular}{c}
    $ \rho$ \\ $\rho u$ \\ $\rho E $
   \end{tabular}
    \right] ,\ 
    &N=\left[
    \begin{tabular}{c}
    $ \rho u$ \\ $p+\rho u^2$ \\ $\rho uH $
   \end{tabular}
    \right], \nonumber
\end{align}
where the operator $\langle \phi \rangle = \phi_1 - \phi_0$ represents the jump of the quantity $\phi$ across the shock front. Assuming the shock front propagates in a still enviroment ($u_0=0$) and by taking the density of the post-shock region, i.e., $\rho_1$, as the free parameter, the post-shock properties for the van der Waals fluid in terms of non-dimensional quantities are obtained as \cite{zhao2011admissible}
\begin{align}
M_0 &= \frac{1}{\tilde{c}_{s0}}\sqrt{6\tilde{\rho}_1\frac{\tilde{p}_0(1+\delta)+\tilde{\rho}_0\tilde{\rho}_1(\tilde{\rho}_0+\tilde{\rho}_1+3\delta-3)}{\tilde{\rho}_0(2\tilde{\rho}_0(3-\tilde{\rho}_1)+3\delta(\tilde{\rho}_0-\tilde{\rho}_1))}}\\
    \tilde{u}_1 &=\tilde{\varsigma}_{1}M_0\frac{\tilde{\rho}_1-\tilde{\rho}_0}{\tilde{\rho}_1}\\
    \tilde{p}_1 &=\tilde{p}_0+\tilde{\varsigma}_{0}^2 M_0^2 \frac{\tilde{\rho}_0(\tilde{\rho}_1-\tilde{\rho}_0)}{\tilde{\rho}_1}
\end{align}
where $(\tilde{u},\tilde{\varsigma})=(u,\varsigma)/\sqrt{p_{cr}v_{cr}}$, $\tilde{p}=p/p_{cr}$ and $\tilde{\rho}=\rho / \rho_{cr}$.

\subsection{SHOCK TUBE}
\begin{figure}
    \centering
    \includegraphics[width=8.6cm]{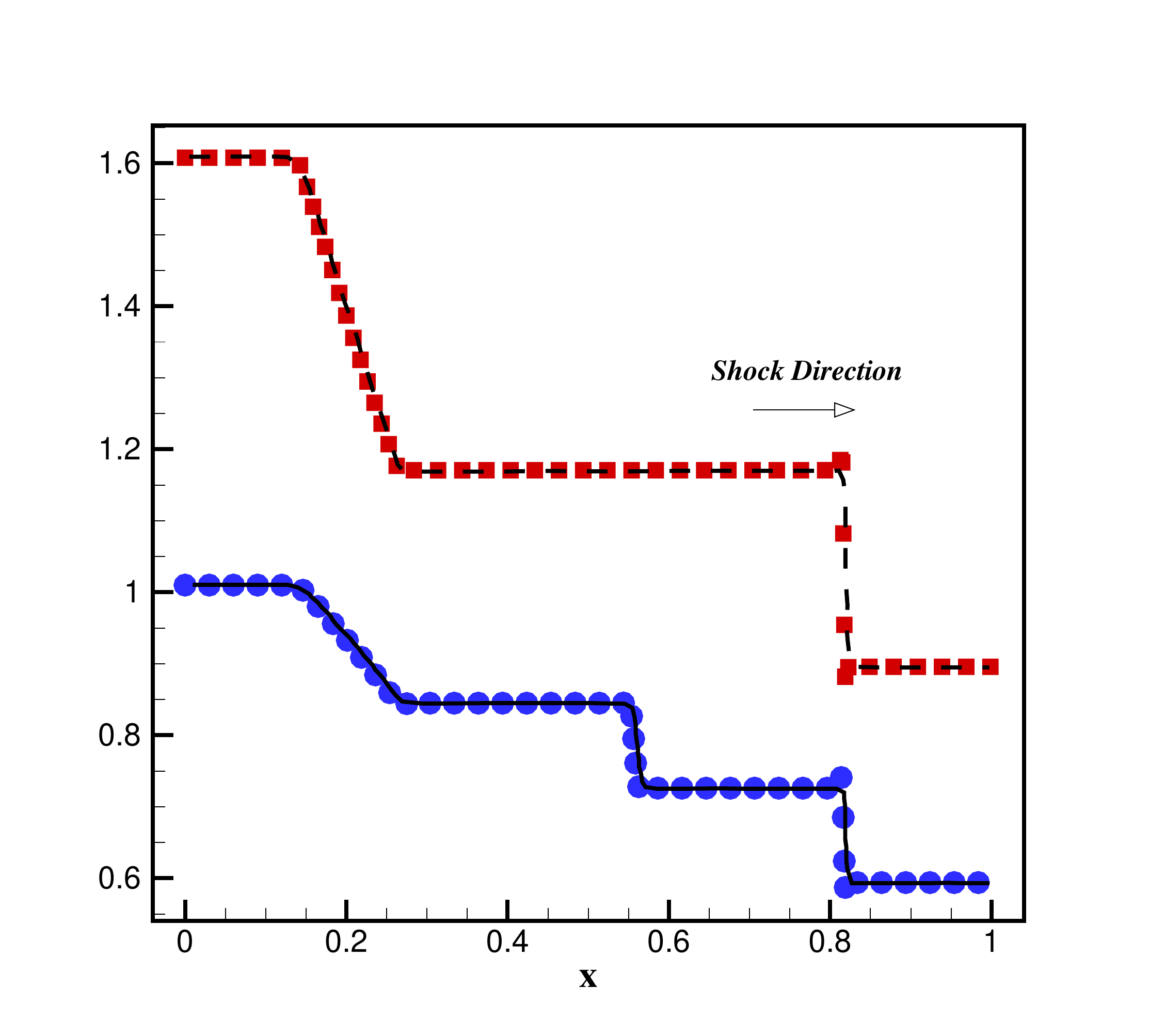}
    \caption{Simulation of shock-tube problem case \Romannum{1} at $t^*=0.2$. Line: density \cite{guardone2002roe}. Dashed: pressure \cite{guardone2002roe}. Symbols: present}
    \label{fig SM shock-tube case1}
\end{figure}

\begin{figure}
    \centering
    \includegraphics[width=8.6cm]{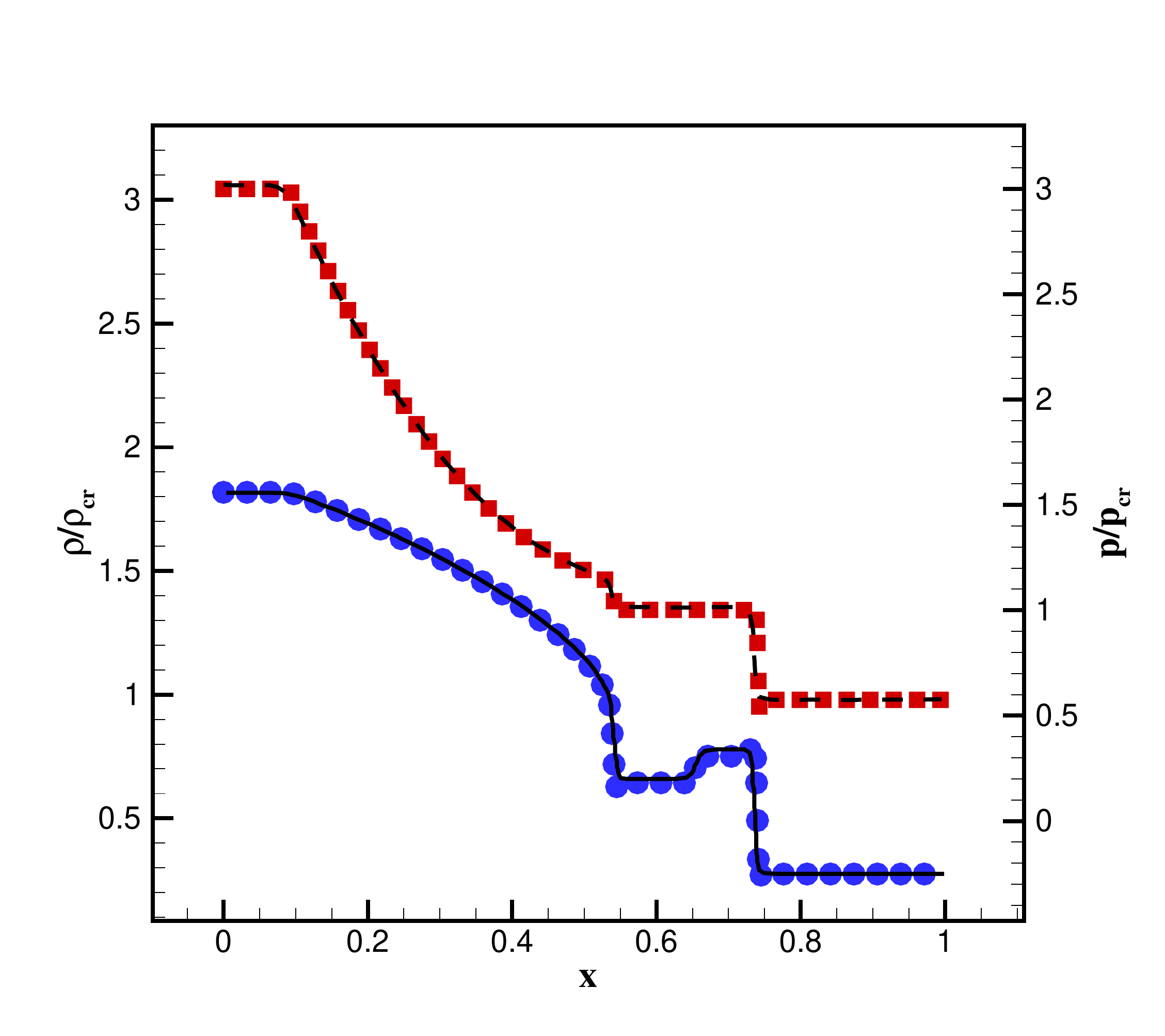}
    \caption{Simulation of shock-tube problem case \Romannum{2} at $t^*=0.15$. Line: density \cite{guardone2002roe}. Dashed: pressure \cite{guardone2002roe}. Symbols: present}
    \label{fig SM shock-tube case2}
\end{figure}

For $\delta\leqslant\delta_{BZT}=0.06$ there exists a region near the saturated vapor, where non-classical behavior may be observed \cite{zhao2011admissible, guardone2002roe}. As mentioned in the main text, this can be characterized by the so called fundamental derivative $\Gamma$. Following the definition of $\Gamma$ and considering the van der Waals fluid, one can derive
\begin{equation}
    \Gamma(p,v) = \frac{(\delta+1)(\delta+2)\frac{p+a/v^2}{(v-b)^2}-\frac{6a}{v^4}}{2(\delta+1)\frac{p+a/v^2}{v(v-b)}-\frac{4a}{v^4}}
\end{equation}
Depending on the sign of $\Gamma$ and whether the $\Gamma=0$ line is crossed during the evolution of the fluid or not, different outcomes may appear \cite{zamfirescu2008admissibility}. According to Ref. \cite{guardone2002roe} two more shock-tube cases are considered here to assess the range of validity of the scheme. The conditions on the left and right sides of the shock tube toghether with the value of $\delta$ corresponding to each case are listed in Table \ref{tab:table shock tube}.
\begin{table}[b]
\caption{\label{tab:table shock tube}
Shock tube conditions
}
\begin{ruledtabular}
\begin{tabular}{ccccc}
\textrm{Case}&
\textrm{$(p_L,\rho_L)$}&
\textrm{$(p_R,\rho_R)$}&
\textrm{$(\Gamma_L,\Gamma_R)$}&
\textrm{$\delta$}\\
\colrule
\Romannum{1} & (1.6077,1.01) & (0.8957,0.594) & (2.239,1.361) & 0.329\\
\Romannum{2} & (3.00,1.818) & (0.575,0.275) & (4.118,0.703) &0.0125\\
\end{tabular}
\end{ruledtabular}
\end{table}
The shock front is initially located on the half-length of the tube. The results for both cases are illustrated in Fig. \ref{fig SM shock-tube case1} and Fig. \ref{fig SM shock-tube case2} at different times. As expected, no anomalous behavior is observed in the first case since $\delta=0.329>\delta_{BZT}$. The pressure and density field shows a classic compression shock traveling towards the right side of the tube accompanied by a rarefaction wave moving to the opposite side. The second case however, is different in terms of existence of non-classical region near the saturation vapor line in the $p-v$ diagram. Although the values of $\Gamma$ for both the left and right parts of the tube is positive, the $\Gamma=0$ line is crossed during the evolution of the flow. This change of sign results in the mixed rarefaction wave composed of a rarefaction shock connected to a rarefaction fan. Meanwhile, a compression shock is traveling toward the low pressure side. The latter case together with the studied item in the main text compose the typical examples of "non-classical gas dynamics", where we can observe anomalous phenomena such as rarefaction shock waves and mixed rarefaction waves. The results show that the present scheme have successfully recovered the non-classic dynamics of a real gas.


\section{CHAPMAN-ENSKOG EXPANSION}

Here we aim at recovering the macroscopic Navier-Stokes equations from the dynamics of kinetic equations (\ref{eq f streaming}-\ref{eq g collision}) in the main text. To this end, the pertinent equilibrium moments of \textit{f} and \textit{g} populations are required, which are computed as follows:
\begin{align}
    P_{\alpha\beta}^{eq}&=\sum_{i=0}^Q f_i^{eq} v_{i\alpha}v_{i\beta}= \rho u_\alpha u_\beta + p\delta_{\alpha\beta}\\
    Q_{\alpha\beta\gamma}^{eq}&=\sum_{i=0}^Q f_i^{eq} v_{i\alpha}v_{i\beta}v_{i\gamma}=\rho u_\alpha u_\beta u_\gamma + p[u\delta]_{\alpha\beta\gamma}\\
    q_{\alpha}^{eq}&=\sum_{i=0}^Q g_i^{eq} v_{i\alpha}= 2\rho u_\alpha H \label{eq SM heatflux}\\
    R_{\alpha\beta}^{eq}&=\sum_{i=0}^Q g_i^{eq} v_{i\alpha}v_{i\beta}= 2\rho u_\alpha u_\beta \left( H+p/\rho\right) + 2pH\delta_{\alpha\beta}, \label{eq SM highordermoment}
\end{align}
where $[u\delta]_{\alpha\beta\gamma}=u_\alpha \delta_{\beta\gamma}+u_\beta \delta_{\alpha\gamma}+u_\gamma \delta_{\alpha\beta}$ and $H$ is the total enthalpy. Note that equations (\ref{eq SM heatflux}) and (\ref{eq SM highordermoment}) are the desired moments, which the equilibrium \textit{$g_i^{eq}$} is constructed upon them.
First, we introduce the following expansions:
\begin{align}
    &f_i = f_i^{{(0)}}+ \epsilon f_i^{(1)} + \epsilon^2 f_i^{(2)}, \label{eq SM f expansion} \\
    &g_i = g_i^{(0)}+ \epsilon g_i^{(1)} + \epsilon^2 g_i^{(2)}, \label{eq SM g expansion}\\
    &\partial_t = \epsilon \dt + \epsilon^2 \dtt,\label{eq SM time expansion}\\
    &\partial_\alpha = \epsilon\dalpha.\label{eq SM spatial expansion}
\end{align}
Applying the Taylor expansion up to second order and separating the orders of $\epsilon$ results in:
\begin{align}
    &\{f_i^{(0)},g_i^{(0)}\}=\{f_i^{eq},g_i^{eq}\},\\
    &\dt \{f_i^{(0)},g_i^{(0)}\} + v_{i\alpha}\dalpha \{f_i^{(0)},g_i^{(0)}\} = -(\omega/\delta t) \{f_i^{(1)},g_i^{(1)}\}, \label{eq SM chapman firstorder}\\
    &\dtt \{f_i^{(0)},g_i^{(0)}\} + \lp \dt+v_{i\alpha}\dalpha \rp (1-\frac{\omega}{2})\{f_i^{(1)},g_i^{(1)}\} \nonumber\\
    &=-(\omega/\delta t) \{f_i^{(2)},g_i^{(2)}\}.\label{eq SM chapman second-order}
\end{align}
The local conservation of density, momentum and energy imply
\begin{align}
\sum_{i=0}^Q\{ f_i^{(n)},g_i^{(n)}\}&=0, n\geq 1 \label{eq SM condition1},\\
\sum_{i=0}^Q f_i^{(n)}v_{i\alpha}&=0, n\geq 1 \label{eq SM condition2}     
\end{align}
Applying conditions (\ref{eq SM condition1}) and (\ref{eq SM condition2}) on equation (\ref{eq SM chapman firstorder}), we derive the following first order equations,
\begin{align}
    &D_t^{(1)}\rho = -\rho\dalpha u_\alpha, \label{eq SM density-firstorder} \\
    &D_t^{(1)}u_\alpha = -\frac{1}{\rho}\dalpha p, \label{eq SM momentim-firstorder}\\
    &D_t^{(1)}T = -\frac{T}{\rho C_v} \lp \frac{\partial p}{\partial T} \rp_\rho \dalpha u_\alpha, \label{eq SM temperature-firstorder}
\end{align}
where $D_t^{(1)}=\dt+u_\alpha \dalpha$ is the first order total-derivative. Note that, the thermodynamic relations (\ref{eq SM real-gas internal energy}-\ref{eq SM vdW internal energy}) together with equations (\ref{eq SM density-firstorder}) and (\ref{eq SM momentim-firstorder}) have been used in deriving the first-order temperature equation (\ref{eq SM temperature-firstorder}). Subsequently, we can derive a similar equation for pressure considering that $p=p(\rho,T)$. This yields
\begin{equation}
    D_t^{(1)}p = \lp \frac{\partial p}{\partial \rho} \rp_T D_t^{(1)}\rho + 
                 \lp \frac{\partial p}{\partial T} \rp_\rho D_t^{(1)}T  
                 = -\rho \varsigma^2 \dalpha u_\alpha, \label{eq SM pressure-firstorder}
\end{equation}
where $c_s^2$ is given by equation (\ref{eq SM speed of sound formula}).\\
The second order relations are obtained by applying the conditions (\ref{eq SM condition1}) and (\ref{eq SM condition2}) on equation (\ref{eq SM chapman second-order}),
\begin{align}
    &\dtt \rho = 0, \label{eq SM density-second-order} \\
    &\dtt u_\alpha = \frac{1}{\rho} \dbeta \left[ \delta t\lp\frac{1}{\omega}-\frac{1}{2}\rp \lp \dt P_{\alpha\beta}^{eq} + \dgamma Q_{\alpha\beta\gamma}^{eq} \rp\right], \label{eq SM momentum-second-order}\\
    &\dtt T = \frac{1}{2\rho C_v} \Bigg\{ \dalpha \left[ \delta t\lp\frac{1}{\omega}-\frac{1}{2}\rp \lp \dt q_{\alpha}^{eq} + \dbeta R_{\alpha\beta}^{eq} \rp\right] \nonumber\\
    & -2\rho u_\alpha \dtt u_\alpha \Bigg\}. \label{eq SM temperature-second-order}    
\end{align}
Equations (\ref{eq SM density-firstorder}) and (\ref{eq SM density-second-order}) constitute the continuity equation. The non-equilibrium pressure tensor and heat flux in the R.H.S of equations (\ref{eq SM momentum-second-order}) and (\ref{eq SM temperature-second-order}) are evaluated using equations (\ref{eq SM density-firstorder}-\ref{eq SM pressure-firstorder}),
\begin{equation}
    \begin{split}
           \dt P_{\alpha\beta}^{eq} + \dgamma Q_{\alpha\beta\gamma}^{eq} &=p\left(\dbeta u_\alpha
   +\dalpha u_\beta\right)\\
   &+\left(p-\rho \varsigma^2\right)\dgamma u_\gamma \delta_{\alpha\beta}
    \end{split}
\end{equation}
\begin{equation}
    \begin{split}
           \dt q_{\alpha}^{eq} + \dbeta R_{\alpha\beta}^{eq}&=2\left(p-\rho \varsigma^2\right)\dgamma u_\gamma u_\alpha\\
           &+ 2p u_\beta\left(\dbeta u_\alpha+\dalpha u_\beta\right)+2p\dalpha h
    \end{split}
\end{equation}
where $h=e+p/\rho$ is the specific enthalpy.
Finally, summing up the contributions of density, momentum and temperature at the $\epsilon$ and $\epsilon^2$ orders, we get the hydrodynamic limit of the model, which reads
\begin{align}
    &D_t\rho=-\rho\nabla\cdot\U, \label{eq SM density}\\
    &\rho D_t\U = -\nabla p -\nabla\cdot\bm{\tau},\label{eq SM momentum} \\
    &\rho C_v D_t T=-\bm{\tau}:\nabla\U-\rho T \left(\frac{\partial s}{\partial \rho}\right)_T D_t\rho
-\nabla\cdot \bm{q}^{\rm neq},\label{eq SM temperature}
\end{align}
We must comment that the expression of heat flux recovered from the Chapman-Enskog analysis without the correction term in the $g$ population is found as $\bm{q}_{\rm CE}^{\rm neq}=- \mu\nabla h,$ where $h=\e+p/\rho$ is the enthalpy. At the limit of an ideal-gas, this is equivalent to the fourier law $\bm{q}_{\rm ig}^{\rm neq}=-k_{\rm ig}\nabla T $ where $k_{\rm ig}=\mu C^{\rm ig}_p$ and hence the Prandtl number is fixed to ${\rm Pr}=\mu C_p^{\rm ig}/k_{\rm ig}=1$ due to the single relaxation time BGK collision model. However, considering the enthalpy of a real-gas as function of pressure and temperature, we have, $\nabla h =C_p\nabla T +v(1-\beta T)\nabla p$, where $v=1/\rho$ is the specific volume, $\beta =v^{-1} \left(\partial v/\partial T\right)_p$ is the thermal expansion coefficient and $C_p = C_v + Tv \beta\lp{\partial p}/{\partial T}\rp_v $ is the specific heat at constant pressure. While one could eliminate the pressure part of the enthalpy by the correction term and only retain the temperature dependent part, it must be noted that the thermal expansion coefficient \textit{at} the critical point diverges, $\beta\to\infty$ and so does the specific heat $C_p\to \infty$. Hence to recover the well-known Fourier law, the post-collision of the $g$ population is modified by adding $\tilde{g}_t \delta t$ where $\sum{\tilde{g}_i} = -2\nabla\cdot(\mu\nabla h) + 2\nabla\cdot (k\nabla T)$ with all the other moments equal to zero. We can take such a population as,
\begin{equation}
    \tilde{g}_i = {M_0}{W_i}{\lp 1+ \frac{\rho(\U\cdot\ci)^2}{2pT_{\rm L}} -\frac{\rho v_i^2}{2p} +\frac{D}{2} \rp}, \label{eq SM gtilde}
\end{equation}{}
where $M_0 = \sum_i{\tilde{g}_i}$. A similar anomaly in the expression for the heat flux was observed and reported in \cite{palmer2000lattice}.

\subsection*{Including the force }

As mentioned in the main text, the force terms in the kinetic equations represent the interface dynamics. First, we recast the post-collision state of the \textit{f} population in the following form \cite{revisedchapman},
\begin{align}
&f_i^* (\x,t) = f_i (\x,t) + \omega (f_i^{eq}(\rho,\hat{\bm{u}}) - f_i (\x,t)) +\hat{S_i},\\
&\bm{\hat{u}}=\bm{u}+\frac{\bm{F}\delta t}{2\rho},\\
&\hat{S}_i=S_i-\omega\left(f_i^{eq}(\rho,\hat{\bm{u}})-\rho W_i\right),
\end{align}
where $[S_i=f_i^{eq}(\rho,\U+\bm{F}\delta t/\rho)-\rho W_i]$ and $\bm{F}=-\nabla\cdot\bm{K}$. Here we expand the forcing term $\hat{S}^{(1)}_i=\epsilon\hat{S}^{(1)}_i$ in addition to the expansions (\ref{eq SM f expansion},\ref{eq SM time expansion},\ref{eq SM spatial expansion}). Similarly, we get the following relations at the orders of $\epsilon^0,\epsilon^1,\epsilon^2$ respectively,
\begin{align}
    &f_i^{(0)}=f_i^{eq}(\rho,\hat{\U}),\\
    &\dt f_i^{(0)} + v_{i\alpha}\dalpha f_i^{(0)} = -(\omega/\delta t) f_i^{(1)}+\frac{1}{\delta t}\hat{S_i}^{(1)},\label{eq SM chapman first-order force}\\
    &\dtt f_i^{(0)} + \lp \dt+v_{i\alpha}\dalpha \rp (1-\frac{\omega}{2})f_i^{(1)}\nonumber\\
    &+\frac{1}{2}\lp \dt+v_{i\alpha}\dalpha\rp \hat{S}_i^{(1)}
    =-(\omega/\delta t) f_i^{(2)},
    \label{eq SM chapman second-order force}
\end{align}
It is important here to assess the solvability conditions imposed by the local conservations. Considering the moment-invariant property of the transfer matrix between the two gauges $\U$ and $\hat{\U}$, one can easily compute,
\begin{align}
    &\sum_{i=0}^Q f_i^{(0)}=\sum_{i=0}^Q f_i^{eq}(\rho,\hat{\U})=\rho,\\
    &\sum_{i=0}^Q f_i^{(0)} v_{i\alpha}=\sum_{i=0}^Q f_i^{eq}(\rho,\hat{\U})v_{i\alpha}=\rho\hat{u}_\alpha.
\end{align}
This implies that,
\begin{align}
    \sum_{i=0}^Q f_i^{(n)}=0, n\geq 1, \label{eq SM condition_force 1} 
\end{align}
\begin{equation}
    \sum_{i=0}^Q f_i^{(n)}v_{i\alpha}= \Bigg\{
    \begin{tabular}{cc}
    $-\frac{\delta t}{2} F_\alpha^{(1)}$ & ,n=1 \\
    0 &  ,n$>$1
    \end{tabular}
    \label{eq SM condition_force 2}
\end{equation}
According to the definition of $S_i$, the following moments can be computed:
\begin{align}
    \sum_{i=0}^Q \hat{S}_i^{(1)} &=0,\\
    \sum_{i=0}^Q \hat{S}_i^{(1)}v_{i\alpha} &=\delta t (1-\frac{\omega}{2})F_\alpha^{(1)},\\
    \sum_{i=0}^Q \hat{S}_i^{(1)}v_{i\alpha}v_{i\beta} &= \delta t(1-\frac{\omega}{2}) \lp \hat{u}_\alpha F_\beta +\hat{u}_\beta F_\alpha \rp+\frac{\omega \delta t^2F_\alpha F_\beta}{4\rho}.
\end{align}
Similarly, the first order equations of density and momentum are derived by applying the solvability conditions (\ref{eq SM condition_force 1}) and (\ref{eq SM condition_force 2}) on equations (\ref{eq SM chapman first-order force}) and (\ref{eq SM chapman second-order force}),
\begin{align}
    &\hat{D}_t^{(1)}\rho = -\rho\dalpha \hat{u}_\alpha, \label{eq SM density-firstorder-force} \\
    &\hat{D}_t^{(1)}\hat{u}_\alpha = -\frac{1}{\rho}\dalpha p+ \frac{1}{\rho}F^{(1)}_\alpha, \label{eq SM momentim-firstorder-force}
\end{align}
where $\hat{D}_t^{(1)}=\dt+\hat{u}_\alpha \dalpha$. At this point it is necessary to mention that since there is a force added to the momentum equation (in this case the divergence of the Korteweg stress), it should also be considered in the energy equation in terms of the work done by that force. Hence the post-collision of the \textit{g} population becomes,
\begin{align}
g_i^* (\x,t) = g_i (\x,t) + \omega (g_i^{eq} - g_i (\x,t))+\tilde{g}_i \delta t +{\phi_i}\delta t,
\end{align}
such that $\sum_{i=0}^Q \phi_i=2\hat{u}_\alpha F_\alpha$. The equilibrium moments are modified as,
\begin{align}
    &\sum_{i=0}^Q g_i^{eq}=2\rho \hat{E},\\
    q_{\alpha}^{eq}=&\sum_{i=0}^Q g_i^{eq} v_{i\alpha}= 2\rho \hat{u}_\alpha \hat{H}, \label{eq SM heatflux-force}\\
    R_{\alpha\beta}^{eq}=&\sum_{i=0}^Q g_i^{eq} v_{i\alpha}v_{i\beta}= 2\rho \hat{u}_\alpha \hat{u}_\beta \left( \hat{H}+p/\rho\right) + 2p\hat{H}\delta_{\alpha\beta}, \label{eq SM highordermoment-force}
\end{align}
where $\hat{E}=e+\hat{u}^2/2$ and $\hat{H}=\hat{E}+p/\rho$. With the changes mentioned so far, the first-order equation of temperature is derived as
\begin{align}
    \hat{D}_t^{(1)}T = -\frac{T}{\rho C_v} \lp \frac{\partial p}{\partial T} \rp_\rho \dalpha \hat{u}_\alpha. \label{eq SM temperature-firstorder-force}
\end{align}
Finally, in a similar manner as the case without the force the macroscopic equations are recovered by collecting the equations of density, momentum and temperature at each order,
\begin{align}
    &\hat{D}_t\rho=-\rho\nabla\cdot\hat{\U}, \label{eq SM density-force}\\
    &\rho \hat{D}_t\U = -\nabla p -\nabla\cdot\hat{\bm{\tau}} -\nabla\cdot \bm{K},\label{eq SM momentum-force} \\
    &\rho C_v \hat{D}_t T=-\hat{\bm{\tau}}:\nabla\hat{\U}- T \left(\frac{\partial p}{\partial T}\right)_v \nabla\cdot\hat{\bm{u}}-\nabla\cdot \bm{q}^{\rm neq},\label{eq SM temperature-force}\\
    &\hat{\bm{\tau}}= -\mu \lp \nabla\hat{\U}+\nabla\hat{\U}^{\dagger}-\frac{2}{D}(\nabla\cdot\hat{\bm{u}})\bm{I}\rp-\eta(\nabla\cdot\hat{\U})\bm{I},
\end{align}
It should be noted that the error terms associated with the forcing are not shown here. For instance, as reported in the literature \cite{revisedchapman,lycett2014multiphase,wagner2006thermodynamic} one can show that the error term in the momentum equation appears as $\nabla\cdot(\delta t^2\bm{F}\bm{F}/4\rho)$.

The total energy of the fluid is formulated by $\hat{E}=e \lp T,v \rp+\hat{u}^2/2+E_\lambda$ where $E_\lambda=\kappa|\nabla\rho|^2/2$ is the non-local part corresponding to the excess energy of the interface. The evolution equation for the specific internal energy $e(T,v)$ can be obtained by considering equations (\ref{eq SM real-gas internal energy final},\ref{eq SM density-force},\ref{eq SM temperature-force})
\begin{equation}
    \rho \hat{D}_t e=-p\nabla\cdot\hat{\U}-\hat{\bm{\tau}}:\nabla\hat{\U}-\nabla\cdot \bm{q}^{\rm neq},
    \label{eq SM internal energy revolution}
\end{equation}
From the momentum equation (\ref{eq SM momentum-force}) we get,
\begin{equation}
    \frac{1}{2}\rho\hat{D}_t \hat{u}^2 = -\hat{\U}\cdot\nabla p -\hat{\bm{u}}\cdot\nabla\cdot\hat{\bm{\tau}}-\hat{\bm{u}}\cdot\nabla\cdot\bm{K},
    \label{eq SM kinetic energy revolution}
\end{equation}
and the evolution of the excess energy can be computed using the continuity equation,
\begin{equation}
    \rho\hat{D}_t E_\lambda=-\bm{K}:\nabla\hat{\U}-\nabla\cdot\left(\kappa\rho\nabla\cdot\hat{\U}\nabla\rho\right), \label{eq SM excess energy revolution}
\end{equation}
by summing up the contribution of each three part we get the full conservation equation for the total energy,
\begin{equation}
      \partial_t\left(\rho \hat{E}\right)+\nabla\cdot\left(\rho \hat{E}\hat{\U}+p\hat{\U}+\hat{\bm{\tau}}\cdot\hat{\U}+
      \bm{K}\cdot\hat{\U}+\kappa\rho\nabla\cdot\hat{\U}\nabla\rho+\bm{q^{neq}}\right)=0.
\end{equation}

%

\end{document}